\documentclass{article}

\usepackage[final,nonatbib]{neurips_2019}

\usepackage[utf8]{inputenc} % allow utf-8 input
\usepackage[T1]{fontenc}    % use 8-bit T1 fonts
\usepackage{hyperref}       % hyperlinks
\usepackage{url}            % simple URL typesetting
\usepackage{booktabs}       % professional-quality tables
\usepackage{amsfonts}       % blackboard math symbols
\usepackage{nicefrac}       % compact symbols for 1/2, etc.
\usepackage{microtype}      % microtypography
\usepackage{subcaption}
\usepackage{color}
\usepackage{mathtools}
\usepackage{wrapfig}

\usepackage[numbers]{natbib}

\newcommand{\eg}{\emph{e.g.}}
\newcommand{\ie}{\emph{i.e.}}

\title{Towards Fair Personalization\\ by Avoiding Feedback Loops}

\author{%
  G\"{o}khan \c{C}apan
  \\\texttt{gokhan.capan@boun.edu.tr}
  \And
  \"{O}zge Bozal
  \\\texttt{ozge.bozal@boun.edu.tr}
  \And
  \.{I}lker G\"{u}ndo\u{g}du
  \\\texttt{ilker.gundogdu@boun.edu.tr}
  \And
  Ali Taylan Cemgil
   \\\texttt{taylan.cemgil@boun.edu.tr}
   \\Department of Computer Engineering, Bo\u{g}azi\c{c}i University, Istanbul, Turkey
}

\begin{document}

\maketitle

\begin{abstract}
%This research is motivated to address fairness issues related to personalized recommender systems. 
Self-reinforcing feedback loops are both cause and effect of over and/or under-presentation of some content in interactive recommender systems. This leads to erroneous user preference estimates, namely, overestimation of over-presented content while violating the right to be presented of each alternative, contrary of which we define as a fair system. 
We consider two models that explicitly incorporate, or ignore the systematic and limited exposure to alternatives. 
By simulations, we demonstrate that ignoring the systematic presentations overestimates \emph{promoted} options and underestimates \emph{censored} alternatives. Simply conditioning on the limited exposure is a remedy for these biases.
\end{abstract}

\section{Introduction}\label{sec:intro}
Contemporary digital platforms provide excessive amount of content imposing on us an exhaustive search to identify the options that match our interest. 
Personalized recommender systems step in as a solution by filtering the alternatives.

Recommendations by the personalized interactive system are based on users' feedback (\eg, clicks, purchases, ratings) to what is recommended by the system in the first place. 
This mechanism has a major impact on the users' future choices, leading to biased \citep{liangcausal, deconvolvingfeedbackloops, debiasingfeedbackloop2019} and inconsistent \citep{interactionrecsys} preference estimates. It gives rise to a vicious cycle that over-presented content gets more attention and keeps being over-presented, whereas under-presented content is subject to going unnoticed and being undervalued.
Ignoring this {\it feedback loop} reinforces the system's prior belief on the user's interest; the over-presented are overestimated and the under-presented are underestimated. 

For instance, {\it popular} (i.e. frequently rated) or {\it promoted} content gets a lot of exposure while less popular ones are under-presented in the recommendations \citep{popularitybias2019, controllingpopularitybias2017,  managingpopularitybias2019}.
On the other hand, relatively favorable alternatives may go under-presented, if at all they are presented. This {\it observation bias} is identified as a significant cause of biased preference estimates \citep{yao2017beyond, fairness2018ucsc}.

Another negative impact of ignoring the feedback loops is \emph{bias towards initial choices}, \ie, initially recommended alternatives are favored by the system. Especially when it does not match the user's interest to begin with, the user feedback might be misleading for the recommender system. As a result, the system may lack in adapting itself to the user's actual preferences, namely, under-presents favorable alternatives. 

Over/under-presentation due to biased preference estimates disturbs the fair setting for both users and content providers. Overlooking some of the options that are actually appealing to the user violates the right of each option to be presented \citep{fairmarketplace2018, reranking2018}, on account of the system's misconception that this option is not preferable by the user.
Eventually, the user experience is isolated to a vicinity of only a limited set of alternatives, leading to a ``filter bubble'' \citep{pariserbubble}. An individual's interests might even go extreme by the systematic exposure \citep{degenerate}.

Undesirable consequences of these scenarios raise the question how a fair and responsible recommender system should operate. 
In this paper, we ask whether incorporating the system's presentations in estimating a user's preference achieve a fair recommender system that is free of bias (\ie, where preference estimates can be adjusted to potentially misleading user behavior towards any limited and systematic presentation), novel (\ie, is capable of eventually finding and presenting to the user her favorite options even if she was not aware of them to begin with), responsible (\eg, avoiding {\it censorship} or overestimation of promoted and popular options) and representative (\ie, sufficiently presenting all options over the course of interactions). 

We then give synthetic examples mimicking challenges towards a ``fair" personalization machinery, and demonstrate that explicitly conditioning on systematic exposure ensures robustness to promotion and unfair comparisons, and discovery of initially censored favorites.

\section{Problem Setup and Methodology}\label{sec:setup}
We assume an interactive system where the user chooses an option $k$ from a systematically selected subset of all alternatives, a {\it presentation} $C$. 
Here, $k \in [K]\coloneqq \{1, 2, \cdots, K\}$, and $C \in \mathcal{C}$, where $\mathcal{C}$ denotes the set of all non-empty subsets of $[K]$.
We assume a discrete choice model, an approach that received recent attention in recommender systems literature \citep[\eg][]{liang2018variational, youtube16deep}, where {\it independence of irrelevant alternatives} holds: relative preferability of an option to another is independent of other alternatives included in, or absent from the presentation. The model also conforms with Luce's choice axiom \citep{choice}: i) options admit a total ordering in their probability of being chosen ii) the user chooses an option from a certain finite subset of the elements with probability proportional to the original choice probability of that option, \ie, the user picks option $k$ from $C$ with probability $p(k\mid C) \propto p(k\mid [K])$.

The task of the hypothesized \emph{fair} personalization system is two-fold. {\it The inference task} is to accurately estimate the user's overall preferences from her choices (\eg, clicks) which are limited to systematically presented subsets of alternatives. Incorporation of presentation in inference is essential to successfully avoid biased or inconsistent estimates \citep{liangcausal, interactionrecsys}. {\it The discovery task}---as $K$ is so large in practice---is to assume responsibility for \emph{finding all good items} without overlooking any alternative \citep{konstan2004evaluating}.

Inference can be carried out with any unbiased estimator of preferences from partial feedback as, \eg,  \citet{efficientbt2012} described for pairwise preferences case and \citet{bayesianplackett2009} or \citet{dlmodel} for $L$-wise ($L<K$) preferences. The discovery task can be handled with a presentation mechanism that explores ``novel'' items, and eventually discovers and composes presentations with the estimated top-$L$ alternatives. Such tasks are widely framed as ``bandit problems'' \citep{banditbook}.

We use the Dirichlet-Luce model and the associated presentation mechanism that \citet{dlmodel} describe. 
This discrete choice model presents a Bayesian treatment of modeling preferences (cast as choice probabilities) by explicitly taking systematic exposure into account, thus eliminating feedback loops. 
The model is shown to approach overall preferences from a {\it presentation-choice} data set
and ensures \emph{independence of unexplored alternatives}---it does not impose negative bias towards never-presented options.
The conjugate Bayesian treatment leads to a straightforward \emph{online learning to present} algorithm based on Thompson sampling \citep{thompson33}, and is shown to achieve superior performance in various online preference learning simulations.
When the systematic presentations are naively ignored, \ie, the user is assumed to pick an option $k$ with probability $p(k\mid [K])$ regardless of the presentation, Dirichlet-Luce reduces to the Dirichlet-Multinomial model.

At this point, we need to clarify some aspects of the assumed setup.  We consider a simplistic single user-system interaction scenario. But as a probabilistic building block, the setup can be reused in a more complex, collaborative recommender system. We ignore the {\it position bias}, \ie, the choice probability being dependent on the position of the option within the presentation \citep{positionbias}. We assume that the user is able to review the shortlisted alternatives and make a choice based on latent preferences. We ignore that the repeated and systematic exposure might actually change users' interests, leading to an \emph{echo chamber}. We refer to \citet{degenerate} for such an analysis, where sufficient conditions that lead to \emph{interest extremes} are provided. It is worth noting that the Dirichlet-Luce model with the associated presentation mechanism at least matches two necessary conditions to avoid such degeneracy points, by naturally allowing a ``growing pool of alternatives'' and due to the randomization inherent in Thompson sampling. We assume that the user picks one of the presented options, although the model can be extended to include an artificial `browse' option corresponding to {\it opting not to choose}. We base this assumption on the ``principle of least effort  \citep{leasteffort},'' \ie, the user would conveniently pick from what is presented to her unless the presented alternatives were too unsatisfactory.
Extensions to the model are conjectured to address some of the limitations as discussed above. The others are irrelevant to our concern, nevertheless, we mention them for clarity. 

We underline that we do not claim that the model is vital to eliminate feedback loops (see, for instance, \citep{deconvolvingfeedbackloops, reranking2018, debiasingfeedbackloop2019} for alternative methods).
Yet we believe, and are concerned with in this paper, that the Dirichlet-Luce model, in comparison to the Dirichlet-Multinomial, provides a test bed to assess whether remedying feedback loops could correct some of the biases commonly encountered in personalized recommender systems.

\section{Simulations}\label{sec:sim}
We design a set of simulations to illustrate that the personalization mechanism of the interactive systems accounting for the systematic exposure, accompanied with a bandit algorithm for online learning to present, addresses biased preference estimates due to over-presentation and under-presentation.

We make a comparison between Dirichlet-Luce and Dirichlet-Multinomial models. The former model infers preferences based on previous choices from systematic presentations that the naive Dirichlet-Multinomial ignores. For both models, we consider a Thompson sampling-based presentation mechanism. Contrary to a greedy mechanism that presents top-$L$ (of $K$) options to the user based on the {\it posterior preference estimate}, Thompson sampling composes a presentation from a {\it posterior preference sample}, thereby achieves exploration of under-presented alternatives. 

We consider an interactive system setup where the user's preferences, represented as choice probabilities $\theta_k = p(k\mid [K])$ $\forall k\in [K]$, are latent. Over the course of interactions, the system presents $\{C_t\mid t=1\cdots T\}$  (succinctly denoted as $C_{1:T}$) and observes choice feedback $k_{1:T}$ (where $k_t \in C_t$). We assume for each interaction, $L=2$ of $K=5$ alternatives are presented to the user, and the user picks one of them. For ease of exposition, we assume $\theta_1>\theta_2>\cdots >\theta_K$ without loss of generality.

\begin{figure}[h]
    \centering
    \includegraphics[width=\linewidth]{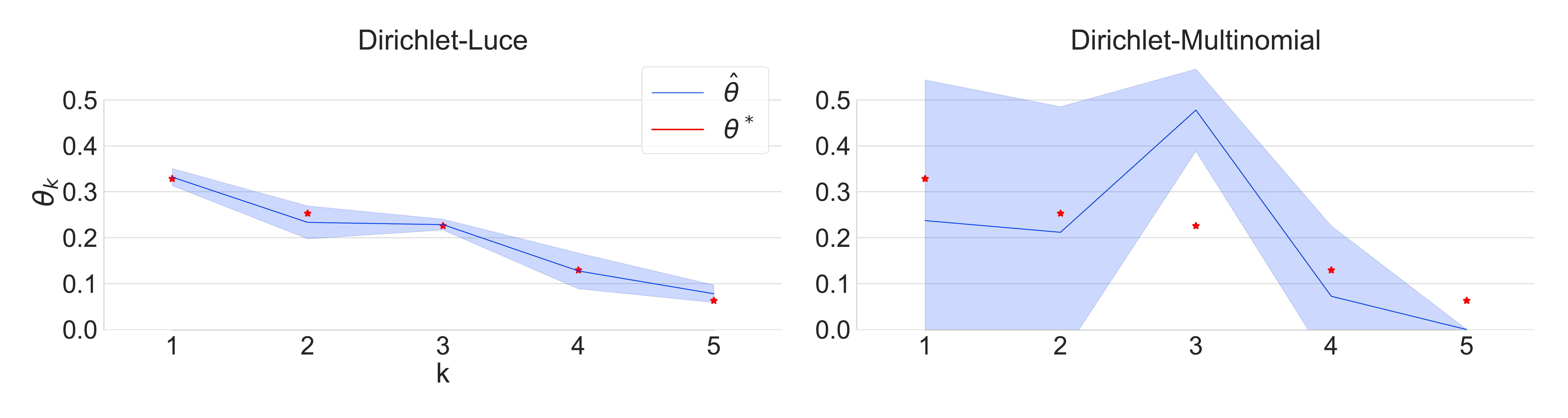}
    \caption{Final (after $T=10000$ interactions) preference estimates ($\hat{\theta}$) averaged over 10 runs when the option 3 is {\it promoted}, \ie, it is included in every presentation. Over-presented option 3 is over-estimated (right), whereas conditioning on presentations fixes this bias (left). Shaded regions denote the standard deviation.}
    \label{fig:promotion}
\end{figure}

Two scenarios are considered to demonstrate how the two models handle potentially biased setups that might lead to underestimation or overestimation of true choice probabilities:

{\bf Robustness to promotion.} We assume a system where a certain option is included in every presentation. A {\it fair} system would \emph{not} overestimate the preference probability of this option. This scenario also subsumes the \emph{popularity bias}, as we artificially include a popular option. Overestimation due to over-presentation can be corrected by conditioning on the presentations, as illustrated in Figure~\ref{fig:promotion}.

\begin{figure}
    \centering
    \begin{subfigure}{\textwidth}
        \includegraphics[width=\linewidth]{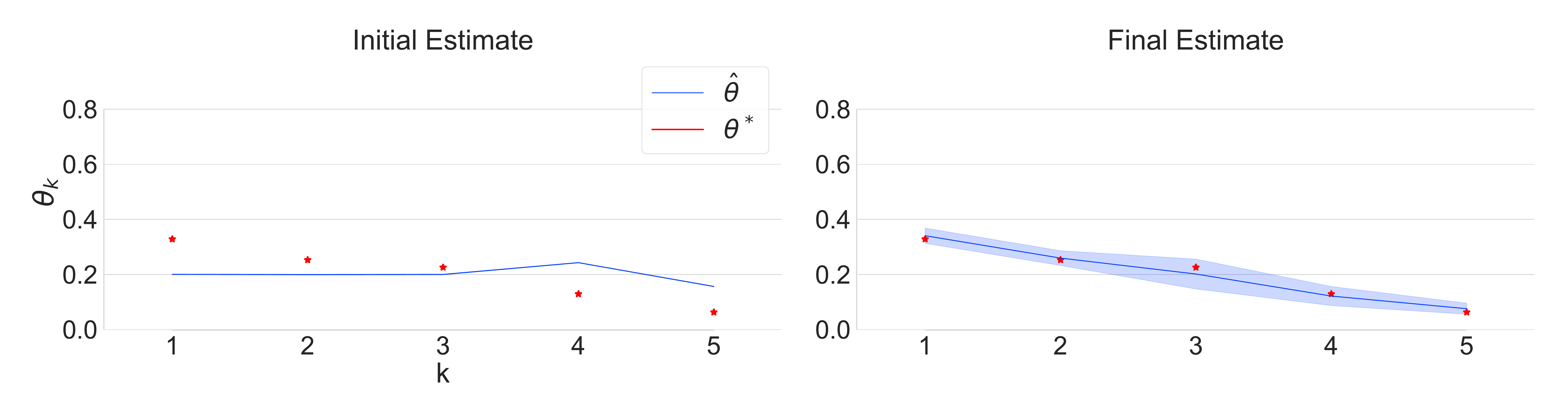}
        \caption{Dirichlet-Luce}
        \label{fig:censoreddl}
    \end{subfigure}
    \begin{subfigure}{\textwidth}
        \includegraphics[width=\linewidth]{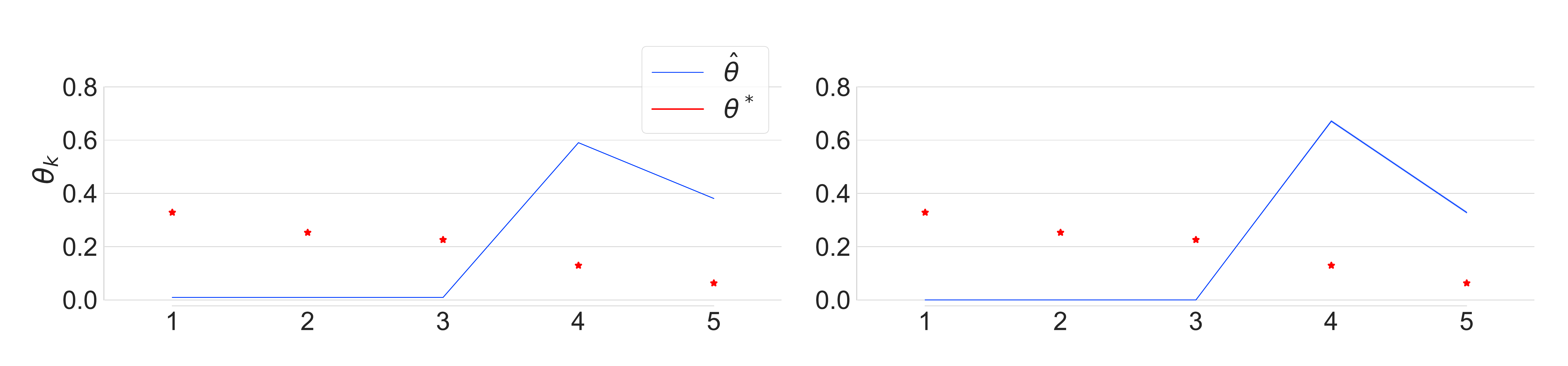}
        \caption{Dirichlet-Multinomial}
        \label{fig:censoreddm}
    \end{subfigure}
    \caption{Preference estimates after 100 choices are made from originally inferior options $\{4,5\}$ (left), and final preference estimates after $T=10000$ interactions (right). Conditioning on presentations does not impose negative bias towards censored options, and the implied personalization model eventually learns to present the best two options (a). On the other hand, over-presented options are overestimated in (b).}
    \label{fig:censored}
\end{figure}

{\bf Discovery of censored favorites.}
Conversely, we assume a system where originally favorable options are initially \emph{censored}, \ie, not included in presentations. A {\it fair} system would correct the bias towards initial choices and give each option equal opportunity to be presented first by not underestimating the choice probabilities of the censored options, and then by eventually learning to present the ``best" options with the discovery mechanism. In Figure~\ref{fig:censored}, we reiterate the illustration in \citep[][Section A.2]{dlmodel}.

It is worth noting that the unbiased results are at the cost of computation. Many quantities of interest for the Dirichlet-Luce model are intractable \citep{dlmodel}, and we resort to Monte Carlo estimates. 

\begin{figure}[h]
    \centering
    \includegraphics[width=\linewidth]{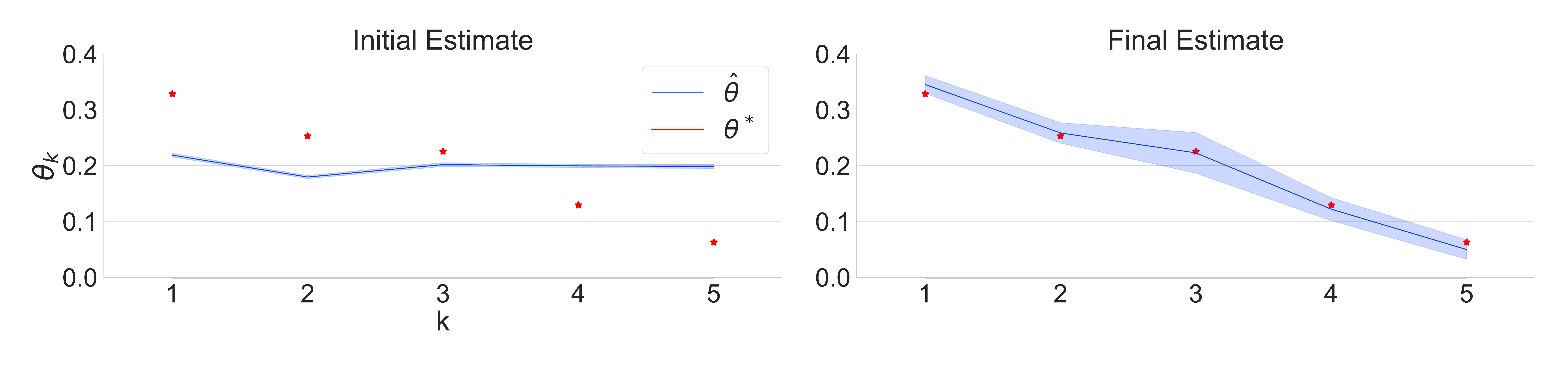}
    \caption{When an initial 100 choices are made from the alternatives $\{1, 2\}$, Dirichlet-Luce initially imposes negative bias towards option $2$ (left). Over the course of interactions, the system captures that option $2$ is still preferable to other, originally inferior options (right).}
    \label{fig:uc}
\end{figure}

{\bf Robustness to unfair comparisons.} The astute reader would suspect that Dirichlet-Luce modeling assumptions would introduce other biases. For instance, repetitive recommendation of a relatively good option along with a more superior one might lead the system to develop a misconception on this option that it is overall inferior. 
The initial misconception is corrected over the course of interactions, as demonstrated in Figure~\ref{fig:uc}.

\section{Discussion}
A personalization system, \eg, a recommender system, ``learns a mechanism from mechanism induced data \citep{mechanism}.'' Ignoring this feedback loop is {\it unfair} to both sides (the user and the content provider) of the interactive system due to overestimation of the over-presented, or underestimation of the under-presented alternatives.
We gave synthetic examples that illustrate {\it promotion} and {\it censorship}, and demonstrated the stark difference between {\it ignoring} and explicitly {\it including} the systematic exposure to the inference mechanism.
As a future work, these experiments can be conducted with multi-user setups and real-world data to check if the personalization mechanism of interest is competent to overcome fairness problems that we investigate. 
\bibliography{bibl}

\end{document}